\documentclass[fleqn,12pt]{wlscirep}
\usepackage[utf8]{inputenc} 
\title{A New Era in the Quest for Dark Matter}

\author[1]{Gianfranco Bertone}
\author[1,2]{Tim M.P. Tait}
\affil[1]{GRAPPA Institute \& Institute of Physics, University of Amsterdam, Science Park 904, 1098 XH Amsterdam, The Netherlands}
\affil[2]{Department of Physics and Astronomy, University of California, Irvine 92697, USA}

\begin{abstract}
There is a growing sense of `crisis' in the dark matter community, due to the absence of evidence for the most popular candidates such as weakly interacting massive particles, axions, and sterile neutrinos, despite the enormous effort that has gone into searching for these particles. Here, we discuss what we have learned about the nature of dark matter from past experiments, and the implications for planned dark matter searches in the next decade. We argue that diversifying the experimental effort, incorporating astronomical surveys and gravitational wave observations, is our best hope to make progress on the dark matter problem.
\end{abstract}
\begin{document}

\flushbottom
\maketitle
\thispagestyle{empty}

\section*{The Fall of Natural WIMPs}

The existence of dark matter has been discussed for more than a century~\cite{Bertone:2016nfn,2017NatAs...1E..59D}. In the 1970s, astronomers and cosmologists have then began to build what is today a compelling body of evidence for this elusive component of the universe, based on a variety of observations that include temperature anisotropies of the Cosmic Microwave Background, baryonic acoustic oscillations, type Ia supernovae, gravitational lensing of galaxy clusters, and rotation curves of galaxies ~\cite{Ade:2015xua,Bertone:2010zza}. 
The Standard Model of particle physics contains no suitable particle to explain these observations, and thus dark matter arguably represents a glimpse of physics beyond the Standard Model (BSM).
Proposed candidates for dark matter span 90 orders of magnitude in mass, ranging from ultra-light bosons, often referred to as ``fuzzy dark matter" \cite{Hui:2016ltb}, to massive primordial black holes, a possibility that has received renewed interest after the LIGO and Virgo detection of gravitational waves from the merger of black holes several tens of times more massive than the Sun \cite{Bird:2016dcv,2017PDU....18..105C}.

The class of dark matter candidates that has attracted the most attention over the past four decades is weakly interacting massive particles (WIMPs). WIMPs appeared for a long time as a perfect dark matter candidate, as new particles at the weak scale would naturally be produced with the right relic abundance in the early universe \cite{Bertone:2004pz}, while at the same time they might alleviate the infamous hierarchy problem\cite{deGouvea:2014xba}
, that has been a main driver of particle physics for roughly four decades\cite{Dine:2015xga}. 
Despite much effort, no particle other than a Standard Model-like Higgs boson has been convincingly detected at the weak scale so far, a circumstance that, as long anticipated \cite{Bertone:2010at}, 
now raises the possibility that natural WIMPs may have been nothing more than an attractive
red herring \cite{Giudice:2017pzm}.

The hierarchy problem is a consequence of the fact that quantum mechanics inevitably mixes up phenomena from all energy scales by allowing virtual particles to participate even in reactions whose energies are far too small to actually produce them.  As a result, low energy quantities, such as the Higgs mass, can potentially receive very large corrections from the virtual influence of much heavier particles.  The influence of heavy particles is particularly pronounced for scalar bosons such as the Higgs, and introduces corrections to the effective Higgs mass proportional to the masses of the virtual heavy states, such that the effective mass is the sum of a fundamental intrinsic value plus the correction terms.  Since it is generally expected that new particles will appear at the Planck energy scale associated with quantum gravity, the observed Higgs mass at the weak scale appears highly unnatural, requiring an incredibly fine-tuned cancellation between the individually much larger intrinsic contribution and the correction terms, such that their sum is the value observed at the Large Hadron Collider.  Natural theories introduce additional particles and symmetries which are arranged in such a way as to cancel off these large corrections amongst each other, protecting the Higgs mass from the influence of heavy mass scales.  

The prototypical natural theory is the minimal supersymmetric (SUSY) standard model, which introduces an additional partner for each Standard Model particle.  In addition, the partners of the electroweak bosons are predicted to be WIMPs, and thus are natural dark matter candidates.  However, most of the parameter space of natural simple supersymmetric models is essentially ruled out \cite{Athron:2017qdc}. Although it is still possible to identify `natural' realizations of SUSY, e.g. in regions of parameter space of the phenomenological Minimal Supersymmetric Model\cite{vanBeekveld:2016hug}, it is undeniable that null searches are constraining larger and larger portions of the parameter space of supersymmetric theories, which begs the question of how much fine-tuning one is willing to accept before giving up the hope to discover SUSY.\cite{Ross:2017kjc}

\section*{Alternatives to Natural WIMPs}

\begin{figure}[t]
\centering
\includegraphics[width=\linewidth]{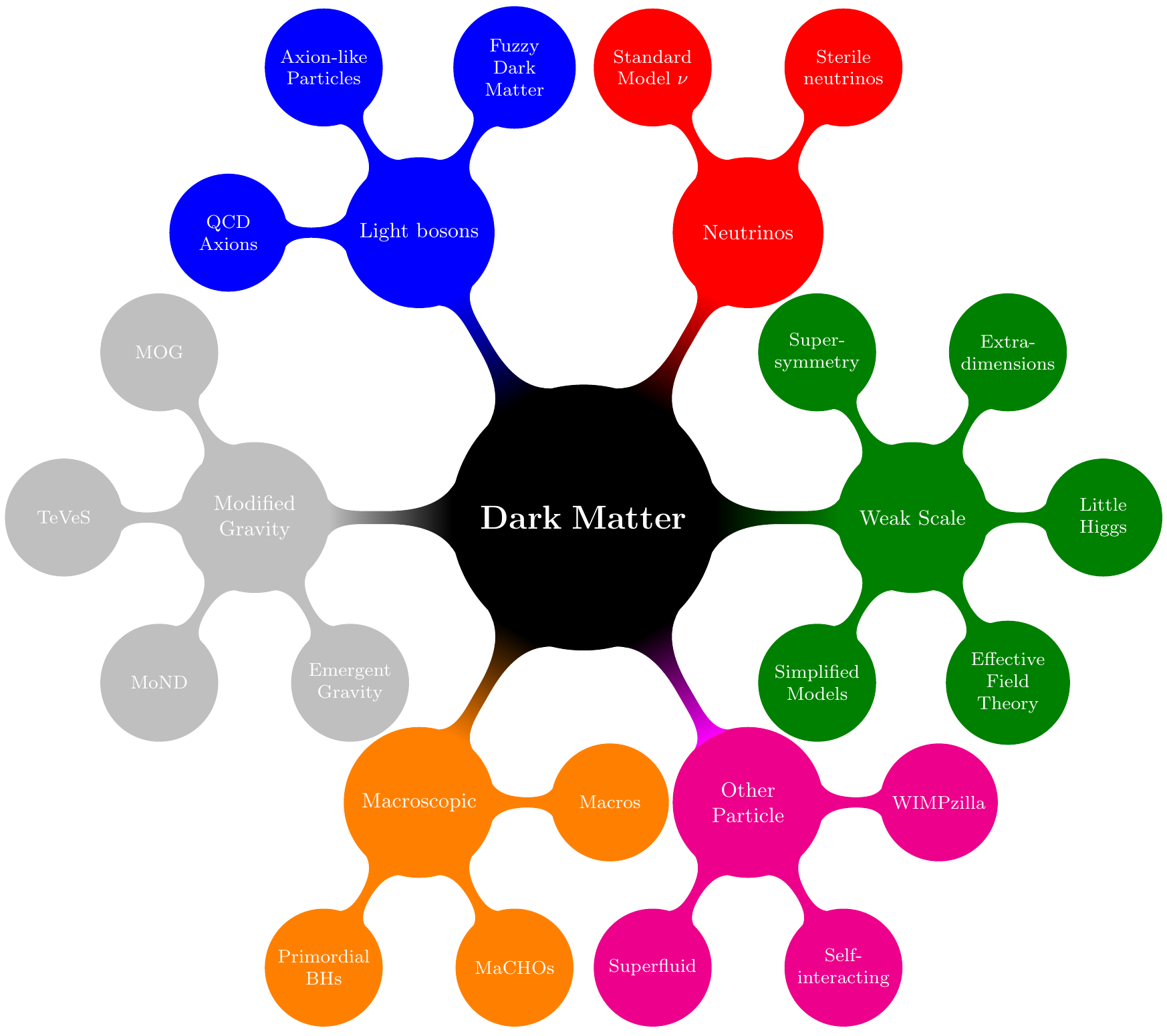}
\caption{Visualization of possible solutions to the dark matter problem.}
\label{fig:venn}
\end{figure}
 
\paragraph{Non-natural WIMPs} 
As a result of the lack of evidence for supersymmetry, {\it naturalness} is beginning to lose its luster as the guiding principle to construct theories of physics beyond the Standard Model. While the shift in emphasis away from WIMPs arising from extensions of the SM that address naturalness is inevitable, WIMPs themselves remain viable dark matter candidates in an appropriate context.  For example, there are types of interactions which lead to indirect and direct signals which are highly suppressed, though such particles remain accessible to the LHC provided their masses are sufficiently small\cite{Goodman:2010ku}.  With naturalness removed as the primary guide to theories of WIMPs, they evolve into a more general class of particles that achieve the appropriate relic density through self-annihilations. 

This wider definition of WIMP -- a circumstance that is already reflected in the adoption of simplified models \cite{Abdallah:2015ter} and effective field theories \cite{Beltran:2010ww} in the presentation of colliders results -- leads to a richer landscape of phenomenology.  For example, the range of WIMP masses expands to encompass masses as low as around 1 MeV, or as high as around 100 TeV.  This wider parameter space demands new kinds of WIMP searches, such as scattering of WIMP-like particles with masses below 1 GeV off of electrons\cite{Essig:2011nj}, or with superconductors\cite{Hochberg:2015pha}, superfluids\cite{Knapen:2016cue}, or Dirac materials\cite{Hochberg:2017wce}.  These light dark matter particles would typically have already been observed if their annihilation cross sections directly into Standard Model particles were large enough to explain their abundance in the Universe.  As a result, viable models typically invoke similarly light dark force carriers, into which the dark matter can annihilate, and which subsequently decay into Standard Model states.  Since they have small masses and must interact at some level with the Standard Model particles, these dark force carriers can be probed by high intensity, low energy accelerators\cite{Essig:2009nc}.
Another complementary avenue is the search for TeV energy gamma rays produced in the annihilation of ultra-heavy dark matter particles with the upcoming gamma-ray Cherenkov Telescope Array (CTA) \cite{Silverwood:2014yza,Acharya:2017ttl}.

\paragraph{Axions} 
Another very popular class of dark matter candidates is that of axions and axion-like candidates. Axions are light ultra-weakly coupled particles which arise as a byproduct in theories which solve the `strong-CP problem.'  The symmetries of the Standard Model of particle physics would allow for the strong nuclear force to include an electric dipole moment (edm) for the neutron, which would represent an asymmetry in the charge distributions of the quarks which make it up.  However, measurements indicate that the neutron edm is about $10^{-10}$ times smaller than naively expected, begging for a dynamical explanation.  The dynamics which would cancel the neutron edm  also produce a new particle: the axion\cite{Abbott:1982af}.

Many constraints exist on axions and axion-like models.  A class of searches typified by the ADMX experiment\cite{Du:2018uak} uses a magnetic field to convert the background of axions on the Earth into an electromagnetic signal, and has successfully excluded a window of axion parameter space with masses around $2~\mu$eV, and future data taking is expected to probe masses up to about $40~\mu$eV.  In addition, there is vigorous theoretical activity exploring new ideas to probe a wider range of axion masses\cite{Kahn:2016aff,Graham:2015ouw,TheMADMAXWorkingGroup:2016hpc}.

\paragraph{Sterile Neutrinos}
Another well-motivated candidate is a sterile neutrino which experiences a diluted form of the weak nuclear force through mixing with the ``ordinary" active neutrinos.  Such particles are a typical ingredient in theories which explain the fact that neutrinos have been experimentally found to be massive, in contrast to the predictions of the Standard Model.  While their residual weak interactions predict that they will ultimately decay, if their mass and mixing are both small enough, the decay may occur slowly enough that they remain in the Universe today as a form of dark matter.  They can be produced in the early Universe through a variety of different physical mechanisms\cite{Shi:1998km,Laine:2008pg,Boyarsky:2009ix,Adhikari:2016bei} with an appropriate abundance.

While the lifetime of a sterile neutrino playing the role of dark matter must be long enough that the vast majority of such particles have not yet decayed, quantum mechanics dictates that some will decay more rapidly, leading to a source of mono-energetic photons with energy close to half of its mass.  In fact, an unidentified emission line at 3.5 keV in the stacked X-ray spectrum of 73 galaxy clusters has prompted the suggestion that it might be a hint of the decay of a sterile neutrino \cite{Bulbul:2014sua}, though debate about the origin of this line is still ongoing \cite{Jeltema:2014qfa}. Future X-ray telescopes such as eRosita, XARM, Athena, and/or Lynx should help to clarify the origin of this emission \cite{Abazajian:2017tcc}, and future accelerator searches such as SHIP will provide a complementary probe of the relevant parameter space.

\paragraph{No stone left unturned.} There is a plethora of other possible explanations for the nature of dark matter (see Figure~\ref{fig:venn} for a diagrammatic representation), from fuzzy dark matter ($10^{-22}$ eV) to gravitationally produced WIMPzillas\cite{Kolb:1998ki}, and from superfluid dark matter\cite{Berezhiani:2015bqa} to macroscopic objects like Macros (10$^{22}$ -- 10$^{24}$g)\cite{2017arXiv170510361K} and primordial black holes (10 $M_{\odot}$).  In light of this situation, the new guiding principle should be ``no stone left unturned": we should look for dark matter not only where theoretical prejudice dictates that we ``must", but wherever we can.  Casting a wider theoretical net offers the possibility to discover new classes of dark matter candidates and new experimental opportunities to search for them, and also helps assemble a ``composite image" of everything we currently know about the space of possibilities consistent with measurements to date. 
\section*{Probing the Nature of Dark Matter with Astronomical Observations}

\paragraph{Departures from LCDM.} 

Given the current absence of evidence for dark matter particles from laboratory experiments, it is of utmost importance to extract as much information as possible from astronomical observations. Dark matter couplings other than gravity with itself or with standard model particle, or a non-negligible velocity dispersion, could lead in principle to measurable differences between observations and LCDM predictions~\cite{Buckley:2017ijx}.  
It is in general important to search for ``cracks" in the LCDM model, by carefully testing its underlying assumptions and observational predictions. An intriguing example is the discrepancy at the 3.7 $\sigma$ level between cosmological \cite{Ade:2015xua} and local measurements of the Hubble constant\cite{2018arXiv180101120R}. 

We stress that great attention has to be paid not to mistake systematic errors in observations, or mismodelling of specific physical processes, for failures of the underlying LCDM model. It is perhaps not a surprise in this sense that most of the claimed problems of standard cosmology, such as the cusp-core, too-big-too-fail, and missing satellites problems~\cite{Buckley:2017ijx}, arise in the deeply non-linear regime. Model predictions are in this case based on numerical simulations that encode complex processes like stellar formation, and supernova and black holes feedback, by means of an effective ``sub-grid" description\cite{2012AnP...524..507F}, which is by construction a potential source of systematic errors. This shouldn't of course deter us from extensively testing the predictions of standard cosmology, by exploiting the wealth of information that will arise from upcoming astronomical surveys such as LSST, DESI, Euclid, and WFIRST, while at the same time improving the quality and predictive power of numerical simulations.   

\paragraph{Self-interactions.} A key property of dark matter that astronomical observations might help disproving is its collisionless nature. Dark matter self-interactions might actually help alleviate claimed tensions between numerical simulations and observations at small cosmological scales\cite{Spergel:1999mh,Tulin:2017ara}. The imprint of dark matter self-interactions can be searched for in a number of ways. First, they can modify the shapes of dark matter halos\cite{Spergel:1999mh}. Self-interactions tend in fact to make the central parts of dark matter halos more spherically symmetric than expected in collisionless scenarios. By comparing the shape of galaxy clusters in numerical simulations with that inferred from lensing and X-ray observations, it is possible to set an upper limit on the velocity-independent, elastic cross-sections of self-interacting dark matter $\sigma /m \gtrsim 1$ cm$^2$ g$^{-1}$~\cite{Brinckmann:2017uve}. Only very recently the first full simulations of galaxy clusters that incorporate both baryonic processes and dark matter self-interactions have been obtained\cite{Robertson:2017mgj}. Although much remains to be understood, it is encouraging that these simulations appear to support the analytical models tying the properties of self-interacting dark matter to the observed distribution of baryons \cite{PhysRevLett.113.021302}. 

The trace of dark-matter self-interactions can also be searched for in merging systems such as cluster mergers and minor infalls \cite{Harvey:2015hha,Robertson:2016qef}. The observables in this case would be the offset between the galaxies and the dark matter (in addition to the offset between dark matter and gas) due to the possible non-collisional nature of dark matter\cite{Randall:2007ph}, and the amount of "sloshing" and "wobbling" of galaxies around the center of the dark matter halo\cite{Harvey:2017afv,Buckley:2017ijx}. As in the case of halo shapes, it is urgent to further investigate with full hydrodynamical simulations the complex interplay between gas cooling, AGN feedback, and dark matter physics, and understand the mapping between the properties of self-interacting dark matter and observables, in preparation of the wealth of observational data that will arise from upcoming astronomical surveys.

\paragraph{Substructures.} A generic key property of dark matter in the standard cosmological model is that is is cold, i.e. non relativistic at the epoch of structure formation, and with a free streaming length much smaller than the size of galaxies. This implies the existence of a large number of sub-dwarf galaxy dark structures in galactic halos. If dark matter is warm, or more in general if its power spectrum is suppressed at small astrophysical scales, then we might discriminate it by probing the actual number of substructures in the Universe. A powerful probe of the power spectrum at small scales is Lyman-alpha forest in the spectra of high-redshift quasars \cite{Narayanan:2000tp}. This technique allows today to set a 2$\sigma$ lower limit on the warm dark matter particle mass of 5.3 keV\cite{Irsic:2017ixq}, and on the mass of fuzzy dark matter particles of 37.5 $\times 10^{-22}$ eV ~\cite{Irsic:2017yje}. New observations with the future high resolution spectrograph of the European Extremely Large Telescope (E-ELT), and with low resolution low signal-to-noise quasar spectra measured by DESI, should allow to substantially improve current bounds thanks to a larger statistical sample and a better determination of the thermal state of the intergalactic medium.

Another intriguing strategy to detect these dark substructures is the search for perturbations induced by sub-dwarf galaxy clumps on cold stellar streams \cite{Yoon2011,Carlberg2012,Bovy2016a}. Thanks to galaxy surveys like Gaia, currently taking data, and LSST, it should be in principle possible to detect impacts induced on stellar streams by subhalos with mass as low as $10^7 ~M_{\odot}$ \cite{Erkal2015a}. By analyzing the power spectrum of the fluctuations of the stellar density in the stream future observations might even allow us to probe subhalos down to $10^5 ~M_{\odot}$\cite{Bovy2016a}.  This method should allow to set stringent constraints on the mass of thermal dark matter relics with LSST data, and possibly yield an actual measurement of the dark matter particle mass if it is in the $\cal{O}$(1) keV range\cite{2018arXiv180404384B}.

A more direct way of detecting dark matter substructures is via gravitational lensing. Although dark matter subhalos are not compact enough to be detectable e.g. with microlensing searches, they can modify the flux ratio of multiply lensed quasars \cite{1998MNRAS.295..587M,2001ApJ...563....9M,2002ApJ...572...25D,2017arXiv171204945G}, and are potentially detectable via gravitational imaging, as a perturbation of  magnified arcs and Einstein rings\cite{Vegetti:2008eg}. On top of lens substructures, low-mass dark matter halos along the line-of-sight can act as perturbers, and dominate the signal by an amount that depends on the lensing configuration and the dark matter properties~\cite{Despali:2017ksx}. This field will soon be revolutionized by upcoming astronomical surveys. The LSST for instance is expected to detect more than 8000 lensed quasars, 13\% of which would be quadruple lenses \cite{2010MNRAS.405.2579O}, which should allow to probe the subhalo mass function below $10^8$ M$_\odot$, while observations in the optical/near infrared with Euclid and the E-ELT, as well as in the radio with ALMA and the global VLBI interferometers should allow to probe the subhalo mass function at high redhsift\cite{Daylan:2017kfh}.

\section*{Gravitational Wave Portal}

\paragraph{Primordial Black Holes.} The detection of gravitational waves~\cite{Abbott:2016blz} has opened up new opportunities to explore the physics of dark matter\cite{2018arXiv180605195B}. First, it has been suggested that the binary black holes whose merger has produced the gravitational waves detected by LIGO might be {\it primordial}, i.e. they might have formed in the very early universe, before Big Bang Nucleosynthesis \cite{Carr:2016drx,Bird:2016dcv,2017PDU....18..105C}. The rate of binary black hole mergers would however be too high if such primordial black holes made up all of the dark matter in the Universe\cite{Sasaki:2016jop,2017PhRvD..96l3523A,2018arXiv180509034K}, a possibility that is also disfavored from a variety of constraints ranging the dynamical heating of dwarf galaxies, to distortions of the Cosmic Microwave Background, and from Supernova lensing to radio and X-ray emission due to the accretion of interstellar gas onto PBHs   ~\cite{Gaggero:2016dpq}. Although the constraints are becoming stringent, it is important to search for these objects even if they represent a subdominant component of dark matter. For instance, if we could discover a population of these objects in the universe, we would know that dark matter is not made of WIMPs, otherwise we should have already detected the annihilation radiation produce by WIMPs around them \cite{Lacki:2010zf}. A number of observations, like the identification of BHs lighter than 1 solar mass, or the existence of BHs at redshift $z \gtrsim 40$ ~\cite{Koushiappas:2017kqm}, may in principle provide strong evidence for the existence of PBHs.  

\paragraph{Constraints on modified gravity.} Since the pioneering work on MOdified Newtonian Dynamics (MOND) in 1982~\cite{Milgrom:1983ca}, numerous attempts have been made (e.g. Modified Gravity (MOG)\cite{Moffat:2005si}, and Emergent Gravity\cite{Verlinde:2016toy}) to get rid of dark matter by modifying Einstein's theory of General Relativity. The success of these efforts however remained limited at most to rotation curves of galaxies, and it is today clear that the only way these theories can be reconciled with observations is by effectively, and very precisely, mimicking the behavior of cold dark matter on cosmological scales. The coincident observation of gravitational waves and electromagnetic radiation from GW170817 \cite{TheLIGOScientific:2017qsa} has allowed to set very stringent constraints on the propagation velocity of gravitational waves. The fact that it does not differ from the speed of light by more than one part in $10^{15}$ severely constrains all theories of modified gravity  in which gravitational waves travel on different geodesics with respect to photons and neutrinos \cite{Boran:2017rdn,Sakstein:2017xjx,Wang:2017rpx}. This has in particular allowed to rule out Bekenstein's Tensor-Vector-Scalar (TeVeS) theory \cite{Bekenstein:2004ne}. 

\paragraph{BH environment.} Interestingly, dark matter might manifest itself as a perturbation in the waveform of binary black holes. If dark matter is made of cold and collisionless particles then their density around black holes will inevitably be higher than on average in the universe, and possibly much higher. Supermassive black holes at the center of galaxies might in particular host dark matter ``spikes"\cite{Gondolo:1999ef}, although dynamical effects such as mergers with other black holes and interactions with stellar cusps might disrupt them\cite{Merritt:2002vj,Bertone:2005hw}. Large dark matter overdensities are possible around Intermediate mass black holes\cite{Bertone:2005xz}, and around primordial black holes\cite{2008ApJ...680..829R}. The presence of dark matter around BHs would modify the dynamics of the merger, and induce a potentially detectable dephasing in the waveform\cite{2018arXiv180605195B}. 

If dark matter is made of ultralight bosons, as in the aforementioned case of fuzzy dark matter, the field “cloud” that forms around black holes with a mass comparable to the Compton wavelength of bosons can be revealed in the gravitational waves signal from single or binary black holes, through direct monochromatic emission, stochastic background or through gaps in the black hole mass-spin Regge plane\cite{Brito:2017zvb,Arvanitaki:2016qwi,2018arXiv180403208B}. Future analyses will allow to further elucidate possible ``environmental" effects due to dark matter particles, and to discriminate among different dark matter models\cite{2018arXiv180605195B}.

\section*{The Future}

In the quest for dark matter, naturalness has been the guiding principle since the dark matter problem was established in the early 1980s. Although the absence of evidence for new physics at the LHC does not rule out completely natural theories, we have argued that a new era in the search for dark matter has begun, the new guiding principle being ``no stone left unturned": from fuzzy dark matter ($10^{-22}$ eV) to primordial black holes (10 $M_{\odot}$), we should look for dark matter wherever we can. 
It is important to exploit to their fullest extent existing experimental facilities, most notably the LHC, whose data might still contain some surprises. And it is important to complete the search for WIMPs with direct detection experiments, until their sensitivity reaches the so-called neutrino floor~\cite{Billard:2013qya}. 

At the same time we believe it is essential to diversify the experimental effort, and to test the properties of dark matter with gravitational waves interferometers and upcoming astronomical surveys, as they can provide complementary information about the nature of dark matter. New opportunities in extracting such information from data arise from the booming field of machine learning, which is currently transforming many aspects of science and society. Machine learning methods have been already applied to a variety of dark matter-related problems, ranging from the identification of WIMPs from particle and astroparticle data~\cite{Bertone:2017adx,Caron:2016hib} to the detection of gravitational lenses\cite{Hezaveh:2017sht}, and from radiation patterns inside jets of quarks and gluons at the LHC~\cite{2017arXiv170904464L} to real-time gravitational waves detection~\cite{George:2017pmj}. In view of this shift of the field of dark matter searches towards a more data-driven approach, we believe it is urgent to fully embrace, and whenever possible to further develop, big data tools that allow to organize in a coherent and systematic way the avalanche of data that will become available in particle physics and astronomy in the next decade. 

\section*{Acknowledgements}

We thank Vitor Cardoso, Daniele Gaggero, David Harvey, Dan Hooper, Bradley Kavanagh, Simona Vegetti, and Matteo Viel for useful comments on the initial version of this manuscript.  The work of TMPT is supported in part by NSF grant PHY-1316792.

\bibliography{sample}

\end{document}